\documentclass[%
 reprint,
superscriptaddress,
 amsmath,amssymb,
 aps,
pra,
]{revtex4-1}

\usepackage{graphicx}
\usepackage{dcolumn}
\usepackage{bm}

\begin{document}

\preprint{APS/123-QED}

\title{Controlling the intensity of  light in large areas at the interfaces of a scattering medium}

\author{Oluwafemi S. Ojambati}\email{o.s.ojambati@utwente.nl}
\affiliation{
Complex Photonic Systems (COPS), MESA+ Institute for Nanotechnology, University of Twente, \\
P.O. Box 217, 7500 AE Enschede, The Netherlands}

\author{John T. Hosmer-Quint}
\affiliation{
Complex Photonic Systems (COPS), MESA+ Institute for Nanotechnology, University of Twente, \\
P.O. Box 217, 7500 AE Enschede, The Netherlands}
\affiliation{%
Department of Physics, Lawrence University, 711 E. Boldt Way,
Appleton, WI 54911, USA}

\author{Klaas-Jan Gorter}
\affiliation{
Complex Photonic Systems (COPS), MESA+ Institute for Nanotechnology, University of Twente, \\
P.O. Box 217, 7500 AE Enschede, The Netherlands}

\author{Allard P. Mosk} \altaffiliation{Current address: Physics of Light in Complex Systems, Debye Institute for Nanomaterials Science, Utrecht University.
}
\affiliation{
Complex Photonic Systems (COPS), MESA+ Institute for Nanotechnology, University of Twente, \\
P.O. Box 217, 7500 AE Enschede, The Netherlands}

\author{Willem L. Vos}
\affiliation {
Complex Photonic Systems (COPS), MESA+ Institute for Nanotechnology, University of Twente, \\
P.O. Box 217, 7500 AE Enschede, The Netherlands}

\date{\today}

\begin{abstract}
The recent advent of wave-shaping methods has demonstrated the focusing of light through and inside even the most strongly scattering materials. 
Typically in wavefront shaping, light is focused in an area with the size of one speckle spot. 
It has been shown that the intensity is not only increased in the target speckle spot, but also in an area outside the optimized speckle spot. 
Consequently, the total transmission is enhanced, even though only the intensity in a single speckle spot is controlled. 
Here, we experimentally study how the intensity enhancement on both interfaces of a scattering medium depends on the optimization area on the transmission side.
We observe that as the optimization radius increases, the enhancement of the total transmitted intensity increases. 
We find a concomitant decrease of the total reflected intensity, which implies an energy redistribution between transmission and reflection channels.
In addition, we find a qualitative evidence of a long-range reflection-transmission correlation.
Our result is useful for efficient light harvesting in solar cells, multi-channel quantum secure communications, imaging, and complex beam delivery through a scattering medium.
\end{abstract}

\pacs{Valid PACS appear here}
\maketitle


\section{Introduction}
\label{S:introduction}

Wave interference in disordered scattering media results in speckles through the coherent addition of multiple waves, which are independent and have random amplitudes and phases~\cite{Goodman2007speckle}.
Between these interfering waves, there exist short-, long-, and even infinite-range correlations~\cite{Shapiro1986PRL,Cwilich1987PRL,feng1988PRL,Zhu1991PRA,Berkovits1994PhysRep,scheffold1998PRL,vanRossum1999RevModPhys,Shapiro1999PRL,BirowosutoPRL2010}.
These correlations have provided enriching information about mesoscopic transport, as well as a deeper understanding of fundamental phenomena such as enhanced backscattering~\cite{Albada1985PRL,Wolf1985PRL,Mueller2001PRA} and Anderson localization~\cite{Anderson1958PRL,conti2007PRA}.

In 1990, using speckle correlations, Freund predicted that an opaque scattering medium can be used as a lens and other optical elements by designing an appropriate incident wavefront~\cite{freund1990PhysicaA}. 
Only recently, this prediction was confirmed by the advent of innovative wave-shaping methods such as 
wavefront shaping~\cite{vellekoop2007OptLett,vellekoop2008PRL, popoff2014PRL,Davy2012PRB,mosk2012NatPhoton,vellekoop2015OptExp}, time reversal~ \cite{Derode1995PRL,lerosey2004PRL,lerosey2007Science}, phase conjugation~\cite{leith1966JOSA,Dowling1992JourAcouAm,yaqoob2008NatPhoton}, and transmission-matrix-based control~\cite{kim2012NatPhoton,Popoff2010PRL,popoff2011NJP}.
In wavefront shaping, an optimization algorithm receives as a feedback the intensity in a target area, typically one speckle spot with an area $ A = \lambda^2/2\pi$.
The algorithm then modifies the spatial phase of the incident field on the scattering medium, such that the intensity in the target spot is maximized.
These wave-shaping methods have led the way for exciting applications such as non-invasive biomedical imaging~\cite{Wang2012NatComm,Si2012NatPhoton,Bertolotti2012Nature}, advanced optics~\cite{Cizmar2010NatPhot,ParkOptExp2012,GuanOptLett2012,Park2012OptLett,SmallOptLett2012,Huisman2015OptExp}, and cryptography and secure communication~\cite{Horstmeyer2013SciReport, Goorden2014Optica}.

In an earlier experiment~\cite{vellekoop2008PRL}, it was observed that there is not only an intensity enhancement in the target speckle spot but also in an area outside the target speckle spot.
Consequently, the total transmission was enhanced, even though only the intensity in a single speckle spot was monitored.
An intuitive explanation for this observation is that there is a redistribution of energy between reflection and transmission channels, since absorption is negligible in the scattering samples.
This observation was confirmed in Ref.~\cite{popoff2014PRL}.
Here, we take a further step by investigating how the enhancement of the total transmission depends on the optimization area.
In the absence of absorption, we expect to observe a concomitant effect in the total reflected intensity.
Moreover, we expect to find the effect of long-range correlations, especially of the form that exists between the reflected and transmitted speckles, as recently predicted in Ref.~\cite{Fayard2015PRA}.
An optimization of the total intensity transmitted through a scattering medium, which is the extreme case of our study, has been performed in Ref.~\cite{popoff2014PRL} although the optimization area was not systematically varied.
The dependency that we seek will give insight to the intensity redistribution between the transmitted and reflected speckles. 
Such a fundamental understanding is useful for applications of wavefront shaping in efficient energy harvesting in solar cells~\cite{levitt1977ApplOpt,polman2012Nature,Si2014APL}, multi-channel quantum secure commun\-ications~\cite{Defiennee2016SciAd,Wolterink2016PRA}, imaging~\cite{yaqoob2008NatPhoton,Vellekoop2010OptLett,Katz2014NatPhoton,Bertolotti2012Nature}, and the delivery of complex beam through a scattering medium~\cite{dickey2014laser}.

In this paper, we experimentally study how the optimized intensity on both interfaces of a scattering medium depends on the optimization area on the transmission side.
We imaged the transmitted intensity onto the chip of a camera, and thus, there is a freedom to control the optimization radius.
We observe that as the optimization radius increases, the enhancement of the total transmitted intensity increases. 
We find a concomitant decrease of the total reflected intensity, which implies that there is a redistribution of intensity from reflection to transmission.
In addition, our result reveals qualitative evidence of the long-range reflection-transmission correlation.

\section{Experimental details}
\label{S:experimentDetails}

\subsection{Experiment set-up}

\begin{figure}[h!] 
\centering
\includegraphics[width = 0.5\textwidth]{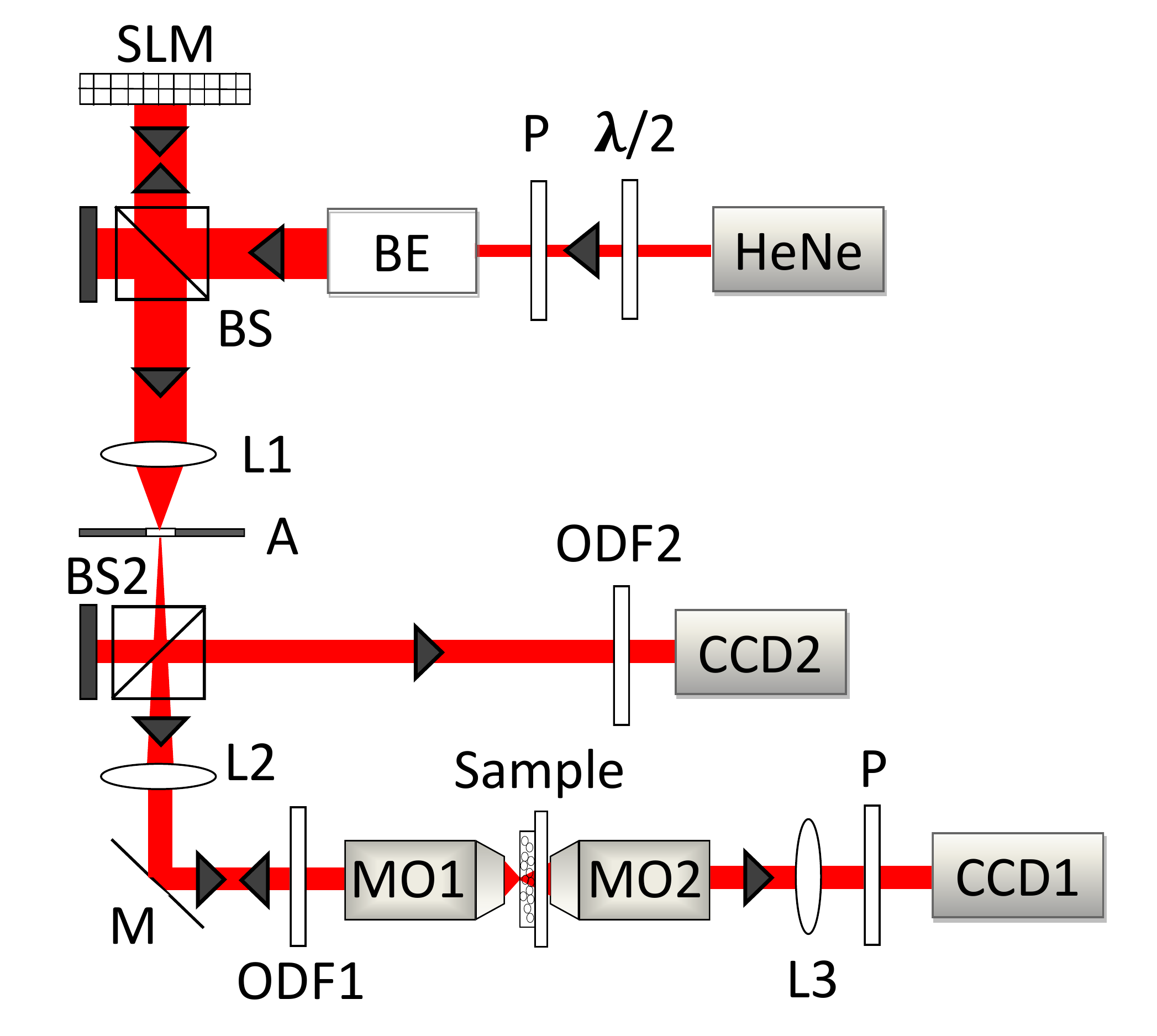}
\caption{Experimental setup. 
A Helium Neon (HeNe) laser beam is expanded and modulated by a spatial light modulator (SLM). 
The light is directed towards the microscope objective (MO1) (numerical aperture NA = 0.95) and then focused onto a multiple scattering sample. 
The sample is made of disordered zinc oxide (ZnO) nanoparticles. 
The intensity transmitted through the sample is imaged onto the chip of a charged-coupled device (CCD) camera (CCD1) through an oil-immersion objective (NA = 1.42) and lens L3.
The reflected intensity is also imaged through a focusing objective and lens L2 and detected by a CCD camera (CCD2). 
P: polarizer, $\lambda/2$: half-wave plate, BE: beam expander, BS: beam splitter, A: aperture, M: mirror.}
\label{fig:setup}
\end{figure}

The experimental set-up is shown in Figure~\ref{fig:setup}. 
The light source is a Helium Neon (HeNe) laser, which emits at a wavelength $\lambda = 632.8$ nm and has an output power of 5 mW.  
A combination of a half-wave plate ($\lambda/2$) and a polarizer (P) controls the incident power and also fixes the polarization of the beam. 
A beam expander with a magnification of $20\times$ expands the beam to fill the active area of the spatial light modulator (SLM). 
The spatial light modulator is a Holoeye Pluto (6010-–NIR-–011), which has 1920 $\times$ 1080 pixels and controls the horizontal polarization. 
A beam splitter (BS) separates the incident and reflected light from the SLM.
The reflected light is focused by a lens L1 (focal length f =  250 mm). 
Aperture A, placed at the focal plane of the lens L1, filters out the higher diffraction orders of the SLM's pixels and transmits only the $0^{th}$ order.
With a telescope consisting of lenses L1 and L2 (f = 250 mm), the SLM is imaged onto the pupil of a microscope objective MO1 (Zeiss: Infinity corrected, 63$\times$, NA = 0.95), which focuses the light onto the surface of the sample. 
The sample is an ensemble of disordered zinc oxide nanoparticles that are spray-painted on top of a glass cover slide.
The sample thickness is 17 $\mu \rm{m} ~ \pm$  0.2 and the transport mean free path $\ell$ of similar samples was reported in Ref.~\cite{PuttenPhdThesis} to be $\ell = 0.6 ~ \mu \rm{m} ~ \pm$ 0.2.
The intensity transmitted through the sample is imaged onto the chip of a charged-coupled device (CCD) camera (CCD1) using a combination of an oil-immersion objective MO2 (Olympus: Infinity corrected, 60$\times$, NA = 1.42) and lens L3 (f = 500 mm). 
The calculated magnification of imaging the back surface of the sample (M1) is $167\times$.  
Similarly, a combination of the focusing objective MO1 and lens L2 images the reflected intensity exiting the front surface of the sample onto the chip of a CCD camera (CCD2).
The calculated magnification on the reflection side is $95\times$.
The cameras CCD1 and CCD2 are both of the same type (AVT Dolphin 145B), with a pixel pitch of 6.45$~\mu $m.
Using the calculated magnifications, the pixel pitches on the front and back surfaces of the sample are $0.068 ~ \mu $m and $0.038 ~ \mu$m respectively.
The optical density filter ODF1 (Thorlabs NE05A), with a measured attenuation factor AF = 0.33, attenuates the incident intensity on the sample, in order to prevent saturation of the cameras. 
The reflected intensity is further attenuated by placing ODF2 (Thorlabs NE10A), with a measured AF = 0.10 in the reflection path. 
The reflected intensity is in total attenuated by a factor of 0.033.

In the set-up of Ref.~\cite{popoff2014PRL}, it is impossible to control the optimization area since the scattering sample was directly attached to a photodetector.
Moreover, as a result of the refractive index contrast (approximately a factor of 2), the detected signal in Ref.~\cite{popoff2014PRL} is limited by significant surface reflections between the scattering sample and photodector. 
With our set-up, we have the freedom to control the optimization area.
A further advantage of our study is that there is no significant surface reflections since there is an index match between the sample substrate and the immersion oil.

\subsection{Experimental procedure and parameters}
In order to optimize multiple speckle spots, we used the partitioning algorithm, which is described in Ref.~\cite{vellekoop2008OptComm} rather than the stepwise sequential algorithm, which is typically used in previous wavefront shaping experiments~\cite{popoff2014PRL,vellekoop2007OptLett,vellekoop2008PRL,AndersonPRA2014,Ojambati2016NJP}. 
We find that the partitioning algorithm outperforms the stepwise sequential and genetic algorithms for optimizing intensity in large areas (see Appendix).
In the optimization, the number of segments used is 5000.
We systematically increased the number of transmission channels by increasing the optimization radius $r_o$. 
The number of open transmission channels $M$ scales linearly with the probed area $A \,(= \pi r_o^2)$ 
\begin{equation}
M = \frac{2\pi A n_{e}^2}{\lambda^2}\frac{\ell}{L} ~ ,
\end{equation}
where $n_e$ is the effective refractive index of the scattering medium~\cite{deBoerPhDThesis,AkbuluttPhdThesis}.
For a specific optimization radius, we repeated the wavefront shaping experiment for  3 to 5 times at a fixed position on the sample. 
As a reference phase pattern, we sent 100 randomly generated patterns, with the same number of segments as the optimized pattern. 
Compared to the optimized pattern, these randomly generated phase patterns have diffraction losses and  a power incident on the sample that is larger by only 5$\%$, which underestimates the intensity enhancement by this amount.

In order to quantify the enhancement $\eta_{\rm{targ}}$ of the total intensity in the target area, we define 
\begin{equation}
\eta_{\rm{targ}} \equiv \frac{P^o_{\rm{targ}}}{\langle  P^u_{\rm{targ}} \rangle}  
\end{equation}
following Refs.~\cite{vellekoop2007OptLett,yilmaz2013BOE}.
$P^o_{\rm{targ}}$ and $P^u_{\rm{targ}}$ are the power in the target area with the optimized and unoptimized patterns, respectively. 
$\langle ~ \rangle$ denotes an ensemble-averaged power over the 100 different random phase patterns.
We also quantified the enhancement $\eta_{\rm{tr},\rm{re}}$ of the total transmitted intensity and the total reflected intensity as 
\begin{equation}
\eta_i \equiv \frac{ P^o_i}{\langle P^u_i \rangle} ~ ,
\end{equation}
where $i = \rm{tr}$ for transmitted light, $i = \rm{re}$ for reflected light, $P^o_{\rm{tr}}$ and $\langle P^u_{\rm{tr}} \rangle$ are the total transmitted power through the sample with the optimized and unoptimized patterns respectively and  $P^o_{\rm{re}}$, and $\langle P^u_{\rm{re}} \rangle$ are the total reflected power through the sample with the optimized and unoptimized patterns respectively. 
The enhancement of the intensity outside the optimization area is quantified as
\begin{equation}
\eta_{\rm{out}} \equiv \frac{ P^o_{\rm{tr}} - P^o_{\rm{targ}} }{\langle P^u_{\rm{tr}} - P^u_{\rm{targ}} \rangle} ~ .
\end{equation}

\section{Results}
\label{s:results}


\subsection{Radial distribution of transmitted intensity}

\begin{figure}[h!]
\centering
\includegraphics[width=0.45\textwidth]{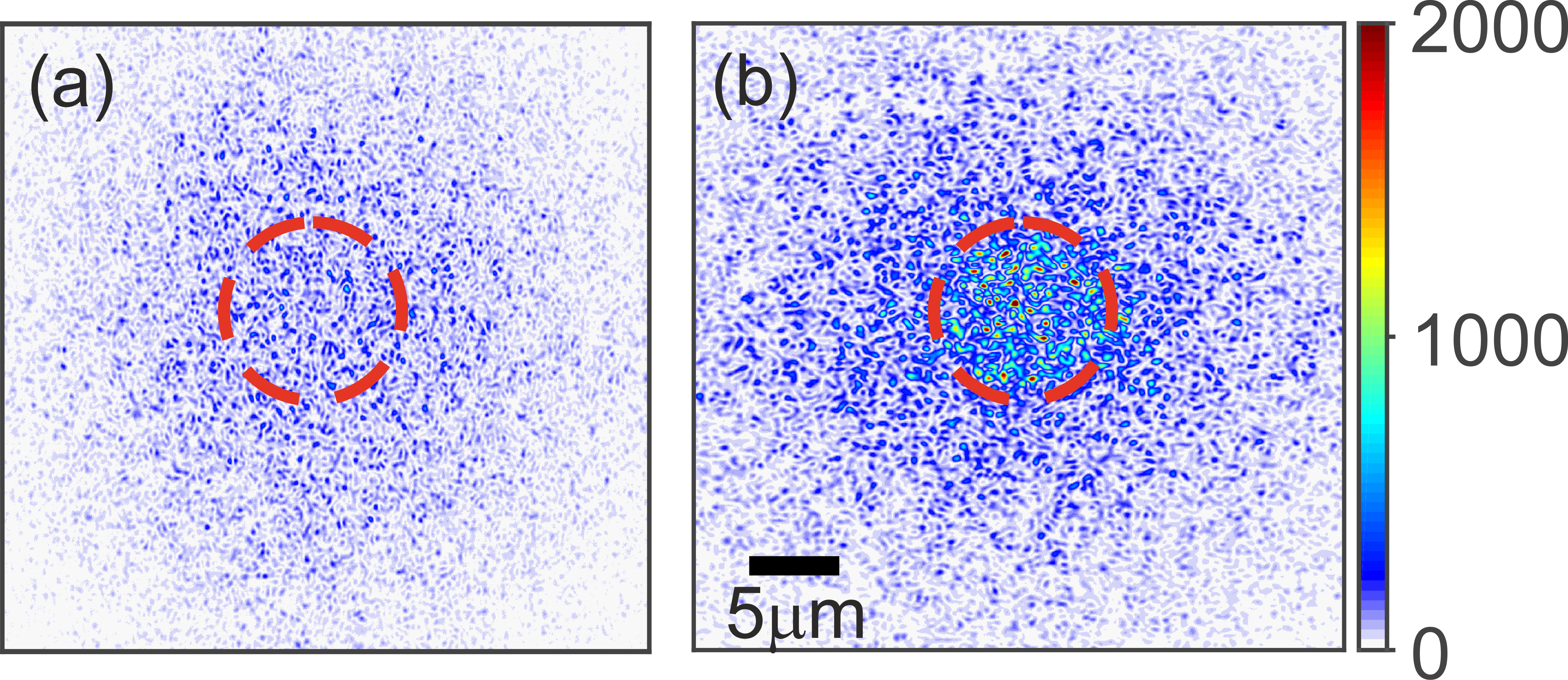}
\caption{Camera images of transmitted intensity at the back surface of a zinc oxide sample. 
In (a) and (b), unoptimized and optimized wavefronts were projected on the spatial light modulator (SLM) respectively. 
The optimization radius $r_o$ = 4.72$~\mu$m, which is indicated by the red dashed circle. }
\label{fig:CCDImages}
\end{figure}

In Figs.~\ref{fig:CCDImages} (a) and (b), we show the CCD camera images of the transmitted intensity for the unoptimized and optimized incident wavefronts, respectively. 
In the wavefront shaping experiment shown in Fig.~\ref{fig:CCDImages} (b), the optimization radius is 4.7$~\mu $m, which corresponds to 121 pixels.
The intensity in the optimization area increases significantly compared to the unoptimized intensity.
The intensity outside the target area increases as well. 
\begin{figure}[h!]
\centering
\includegraphics[width = 0.4 \textwidth]{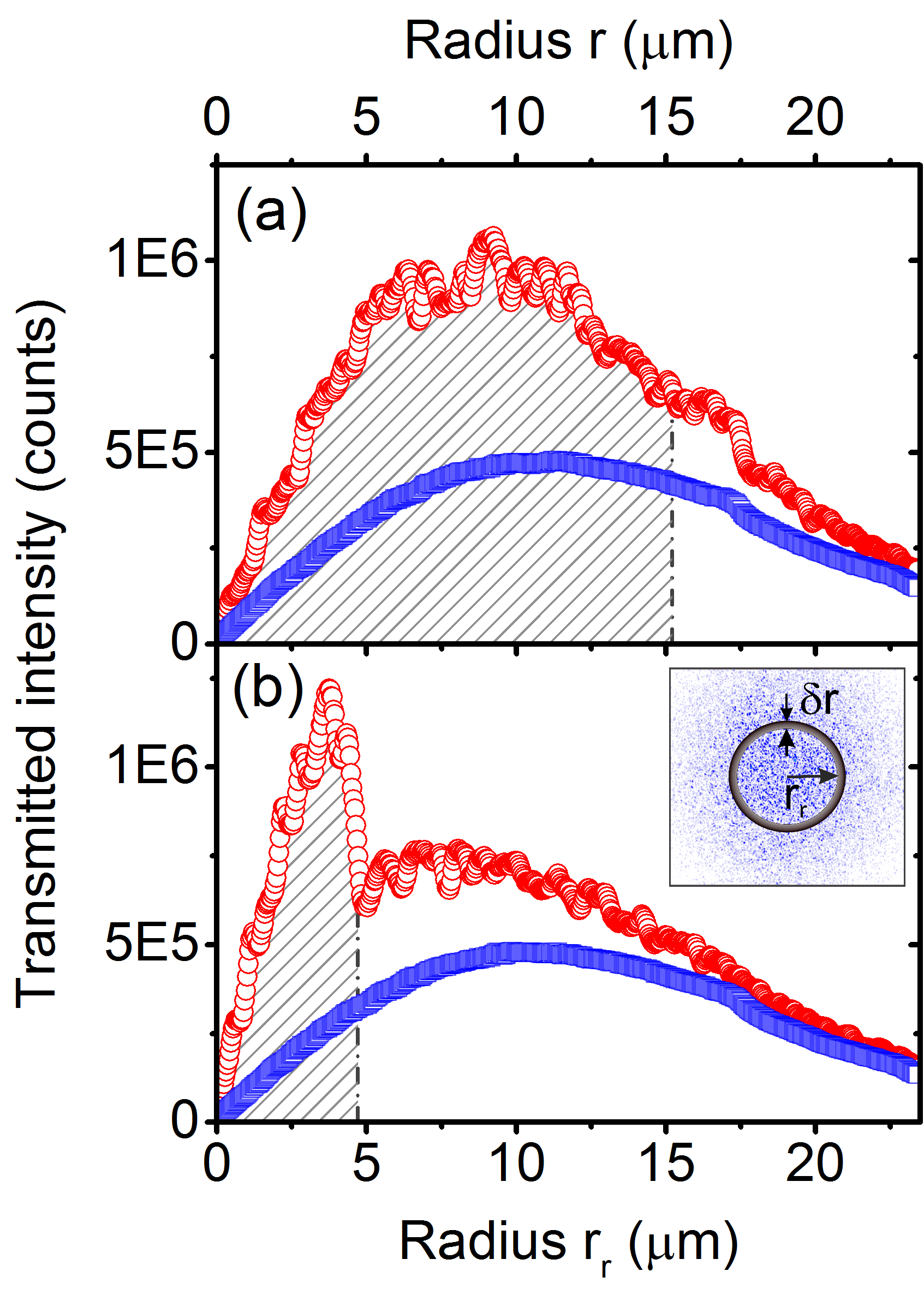}
\caption{Transmitted intensity versus radius $r$. The transmitted intensity is the total intensity within the circumference of a ring, which has an inner radius $r$ and ring width $\delta r$ (see inset).
The optimization radii $r_o$ are (a) $r_o$ = 15.2 $~ \mu $m and (b) $r_o$ = 4.7 $~ \mu $m.
The hatched area under the curves is the optimization area.
The red circles and blue squares are the intensities for the optimized and unoptimized incident wavefronts respectively.
}
\label{fig:TransmtIntensity_ring}
\end{figure}
For a better visualization of the intensity increase, we plot in Fig.~\ref{fig:TransmtIntensity_ring} the radial distributions of the transmitted intensity. 
The radial distribution is obtained by summing the intensities within a ring with a width $\delta r$ and an inner radius of $r_r$ (see inset in Fig.~\ref{fig:TransmtIntensity_ring}).
Angular averaging helps to reduce the intensity fluctuation from the speckle pattern.
There is a significant intensity increase in the optimization area for both optimization radii $r_o$ = 15.2$~\mu $m and $r_o$ = 4.7$~\mu$m. 
This intensity increase is expected since the intensity in the optimization area is the feedback to the partitioning algorithm.
The intensity outside the optimization area remarkably increases as well. 
This intensity increase agrees with the observation in Ref.~\cite{vellekoop2008PRL}, where the intensity outside the optimization area was observed to increase as well. 
We quantify the intensity enhancement inside and outside optimization areas, and the total transmitted intensity in the next sections.

\subsection{Enhancement of the intensity in the optimization area}
\begin{figure}[h!]
\centering
\includegraphics[width = 0.45 \textwidth]{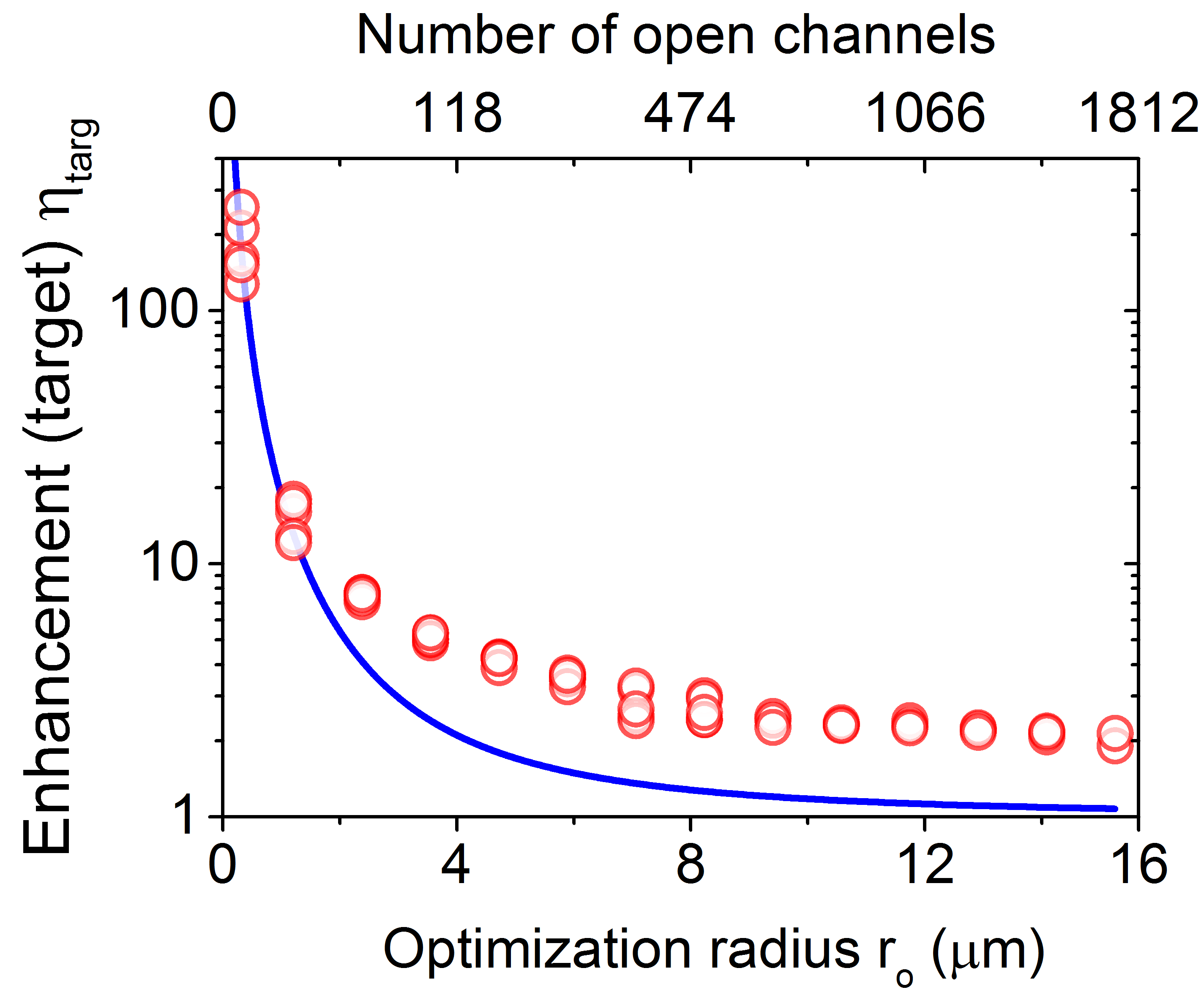}
\caption{Enhancement $\eta_{\rm{targ}}$ in the target area versus optimization radius $r_o$. 
The red circles are the experimental data and the blue curve is an inverse square law fit following the prediction of Ref.~\cite{Vellekoop2008OE}. }
\label{fig:eta_target}
\end{figure}

We plot in Fig.~\ref{fig:eta_target} the intensity enhancement $\eta_{\rm{targ}}$ in the optimization area versus the optimization radius $r_o$.  
We obtained a maximum enhancement of 257$\times$ for an optimization radius of about 0.3$~\mu $m, which corresponds to one speckle spot. 
As the optimization radius increases, $\eta_{\rm{targ}}$ decreases drastically down to 2$\times$ at the largest radius $r_o = $ 15.2$~ \mu$m, which corresponds to 1716 open transmission channels.
In Ref.~{\cite{Vellekoop2008OE}}, the enhancement for a multiple channel optimization was found to be inversely proportional to the number of speckle spots $M$ in the optimization area. 
In the model in ~{\cite{Vellekoop2008OE}}, it is assumed that the optimized intensity is distributed equally to all optimized channels, which are considered statistically independent. 
From the model, the enhancement is expected to depend inversely on the square of the optimization radius $r_o$, 
\begin{equation}
\eta_{\rm{targ}} = \frac{K}{r_o^{2}} + 1~ ,
\label{eq:eta}
\end{equation} 
where $K$ is a constant that depends on the number of effectively controlled input channels on the sample. 
In Fig.~\ref{fig:eta_target}, we show a nonlinear least squares fit to the experimental data using Eq.~\ref{eq:eta} and $K$ is the only adjustable parameter.
Weighing all data points equally, the fit agrees well with the first two optimization radii $r_o$ = 0.3$~\mu $m and 1$~\mu$m that have 1 and 7 transmission channels respectively. 
Strikingly, the fit deviates from the experimental data by about a factor of 2 at large radii. 
This deviation signifies that there is more intensity available in the optimization channels, than that predicted according to Eq.~\ref{eq:eta}, especially at large radii.  

We discuss three possible reasons for the deviation of the theory from the experiment\-al data. 
First, it is known that noise has a significant effect on the single-speckle optimizati\-on~\cite{yilmaz2013BOE}. 
Our wavefront shaping experiments are in the regime where shot noise is much higher than the camera and laser noise that are about 1$\%$ and 2$\%$ respectively.
According to Ref.~\cite{yilmaz2013BOE}, in this shot noise regime the enhancement of a single-speckle optimization is expected to scale linearly with the total intensity in the optimization area $P_{\rm{targ}}$.
Extending this model to the optimization of multiple channels, we derive
\begin{equation}
\eta_{\rm{targ}} = \frac{K P_{\rm{targ}}}{r_o^{2}} + 1 = K^\prime + 1~ ,
\end{equation}
where $K^\prime \equiv K C$, where $C$ is a pre-factor in the relationship $P_{\rm{targ}} = C r_o^2$.
A constant enhancement with radius obviously does not describe our experimental data, hence we reject this hypothesis. 

Second, the observed increased enhancement might be due to intensity redistributed from the speckles outside the optimized area to speckles inside the optimized area.
If this is the case, then the total transmitted intensity would be constant for all optimization radii. 
A third hypothesis is that there is a redistribution of intensity from the reflected speckles to the transmitted speckles. 
In this case, the effect of enhancing the transmitted intensity is expected to be noticeable on reflection as a reduction of the reflected intensity.
We will check these later two hypotheses in the next section.

\subsection{Change of both transmitted and reflected intensities}
\label{ss:eta_total}

\begin{figure}[tb!]
\centering
\includegraphics[width = 0.45 \textwidth]{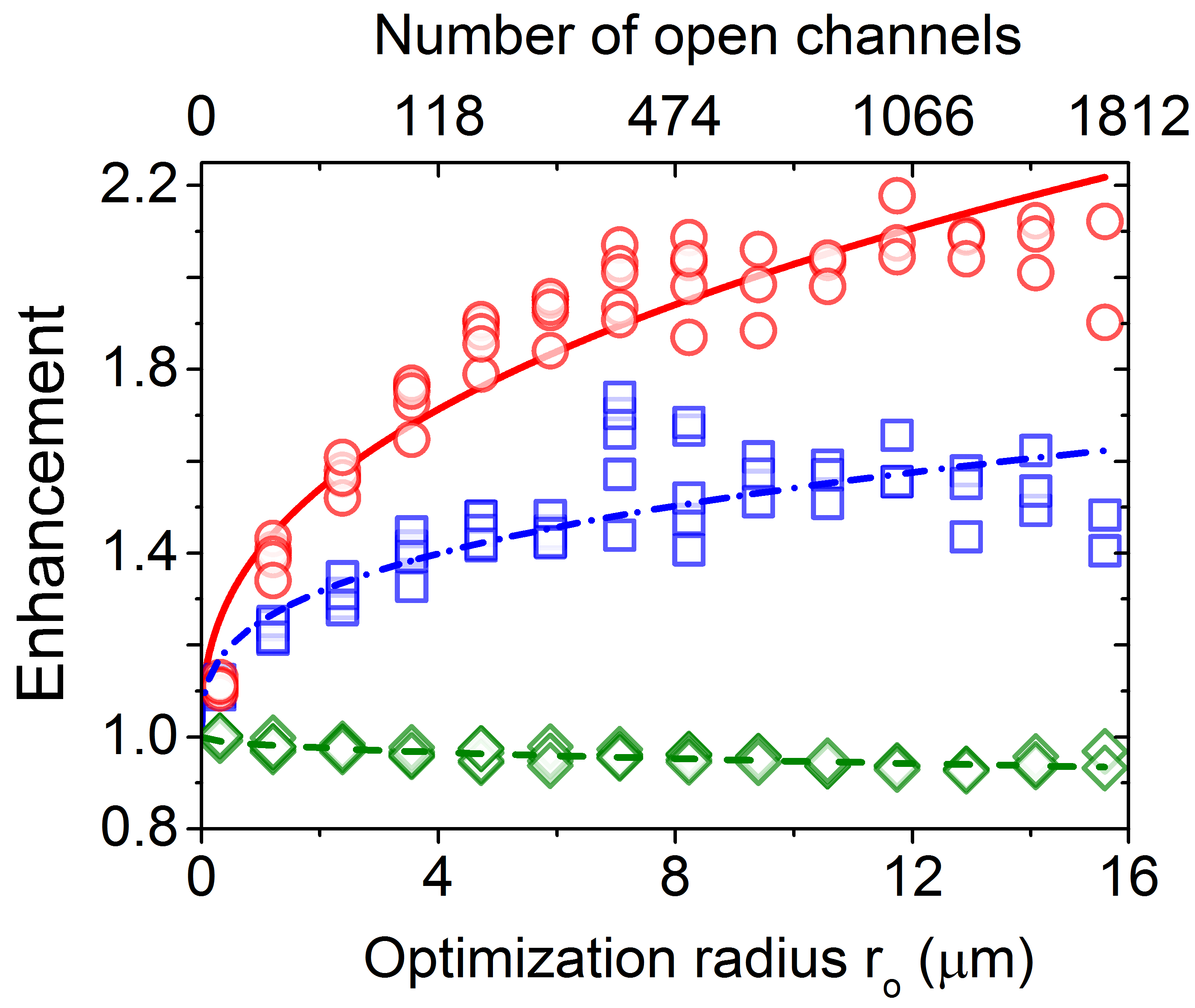}
\caption{Enhancement versus optimization radius $r_o$. 
The red circles are the experimental data of the enhancement of the total transmitted intensity. 
The blue squares are the enhancement of the intensity in the area outside the target area and the green diamonds are the enhancement of the total reflected intensity.
The red solid curve, blue dash-dot curve and the green dash curve are fits to the experimental data using Eq.~\ref{eq:eta_total} and the parameters are given in Table~\ref{table:table1}.}
\label{fig:eta_total}
\end{figure}

The enhancements of the total transmitted intensity $\eta_{\rm{tr}}$, the total reflected intensity $\eta_{\rm{re}}$, and the intensity outside the optimization area $\eta_{\rm{out}}$ versus the optimization radius are shown in Fig.~\ref{fig:eta_total}. 
The enhancement of the total transmitted intensity gradually increases from about 1.1 to 2 at the largest radius of 15.2$~\mu $m, which is close to the $17 ~ \mu $m large size of the detected speckle pattern.
There is also an enhancement of the intensity outside the optimization area and it is about $\eta_{\rm{out}} \approx 1.5$ at large radii.
The enhancements of both the total transmitted intensity and the intensity of area outside the optimization nullifies the second hypothesis. 
In contrast to the transmitted intensity, the enhancement of the total reflected intensity slowly decreases to $\eta_{\rm{re}} \approx 0.93$ as the optimization radius increases. 
The decrease of the $\eta_{\rm{re}}$ is not as rapid as the increase of $\eta_{\rm{tot}}$ because the reflected intensity is about nine times higher than the transmitted intensity. 
Therefore, a large intensity enhancement on transmission corresponds to a small intensity enhancement on reflection.

In order to accurately compare the decrease of $\eta_{\rm{re}}$ with the increase of $\eta_{\rm{tot}}$, we need to know the dependence of both terms on the optimization radius.
The dependence of the enhancements on the optimization radius is unknown and we find that a power-law
\begin{equation}
\eta = \frac{B}{r_o^n} + 1
\label{eq:eta_total}
\end{equation}
describes the experimental data well. 
Here, $B$ and $n$ are adjustable parameters.
The fits to the experimental data are shown in Fig.~\ref{fig:eta_total} and the values of $B$ and $n$ obtained from the fits are shown in Table~\ref{table:table1}. 
We obtained $n = 0.4$ and $n = 0.5$ for the enhancement of the total transmitted and reflected intensities, respectively, and these values are in remarkable mutual agreement. 
\begin{table}[t]
\centering
\begin{tabular}{ |c|c|c| } 
 \hline
  & B & n  \\ 
 \hline \hline
 $\eta_{\rm{tot}}$ 	& 	0.42 & 0.4 \\ 
$\eta_{\rm{re}}$	& 	-0.02 & 0.5 \\ 
$\eta_{\rm{out}}$   & 	0.25 & 0.35\\ 
 \hline
\end{tabular}
 \caption{The values of the adjustable parameters B and n obtained by fitting Eq.~\ref{eq:eta_total} to the experimental data shown in Fig.~\ref{fig:eta_total} for the total transmission enhancement $\eta_{\rm{tot}}$, total reflection enhancement $\eta_{\rm{re}}$ and the enhancement of the intensity outside the target area $\eta_{\rm{out}}$}.
 \label{table:table1}
\end{table}
\begin{figure}[h!]
\centering
\includegraphics[width = 0.45 \textwidth]{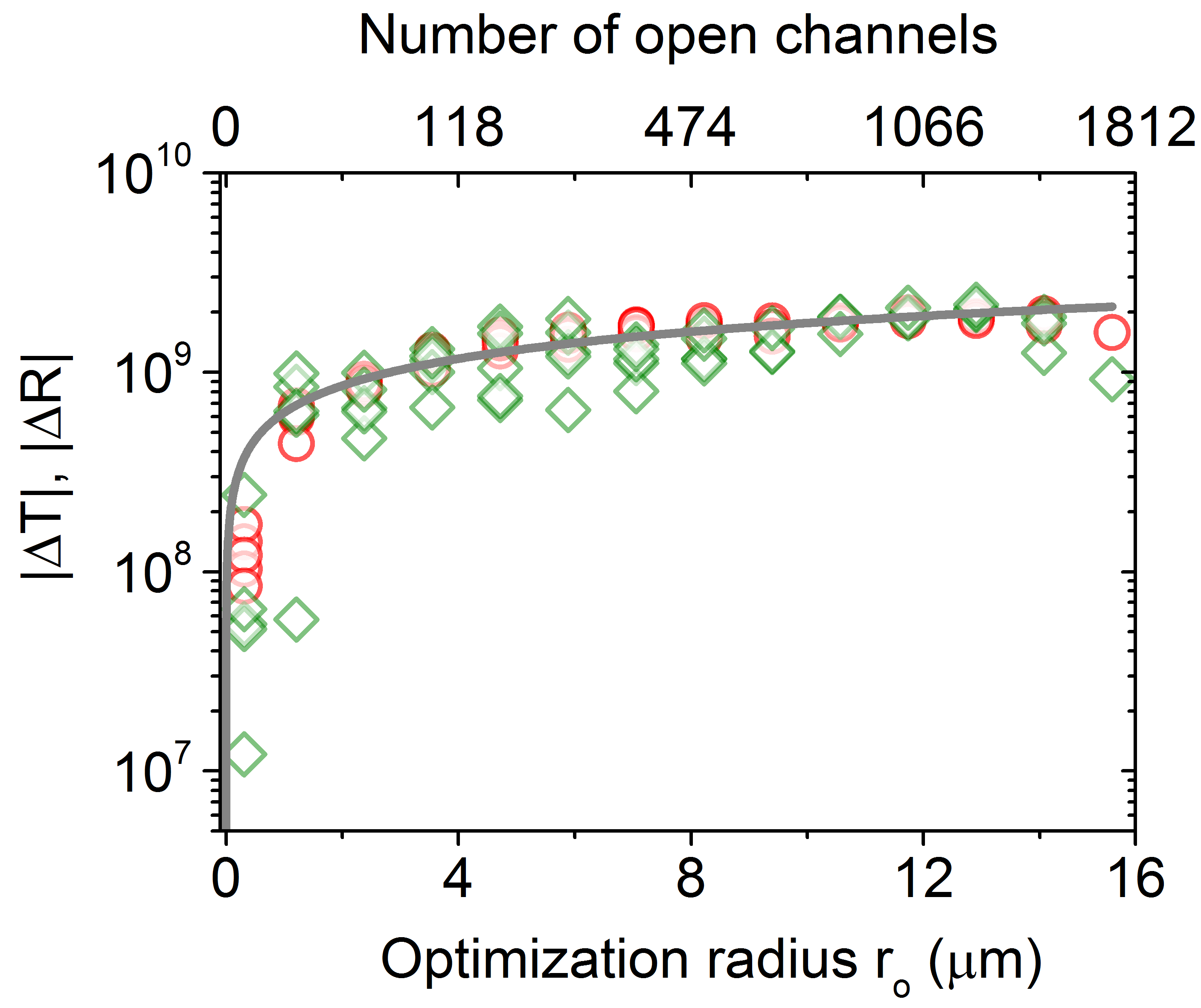}
\caption{Measured absolute change in the total transmitted intensity ($\Delta T$) (red circles) and total reflected intensity ($\Delta R$) (green diamond) versus optimization radius $r_o$.
The solid gray curve is a fit using Eq.~\ref{eq:eta_total}, with $B = 6.3\times 10^8$ and $n = 0.45$.}
\label{fig:intensityChange}
\end{figure}
In Fig.~\ref{fig:intensityChange}, we plot the absolute changes in the transmitted and the reflected intensities after accounting for the attenuation factor of the ND filters.
The absolute changes in the transmitted and reflected intensities both collapse to the same curve, modeled with $n = 0.45$.
Therefore, the enhancement of the transmitted intensity corresponds to a decrease in the reflected intensity. 
This validates the third hypothesis that there is a redistribution of intensity from the reflection speckles to the transmission speckles of the scattering medium.

\subsection{Radial distribution of reflected intensity}

\begin{figure}[h!]
\centering
\includegraphics[width = 0.45 \textwidth]{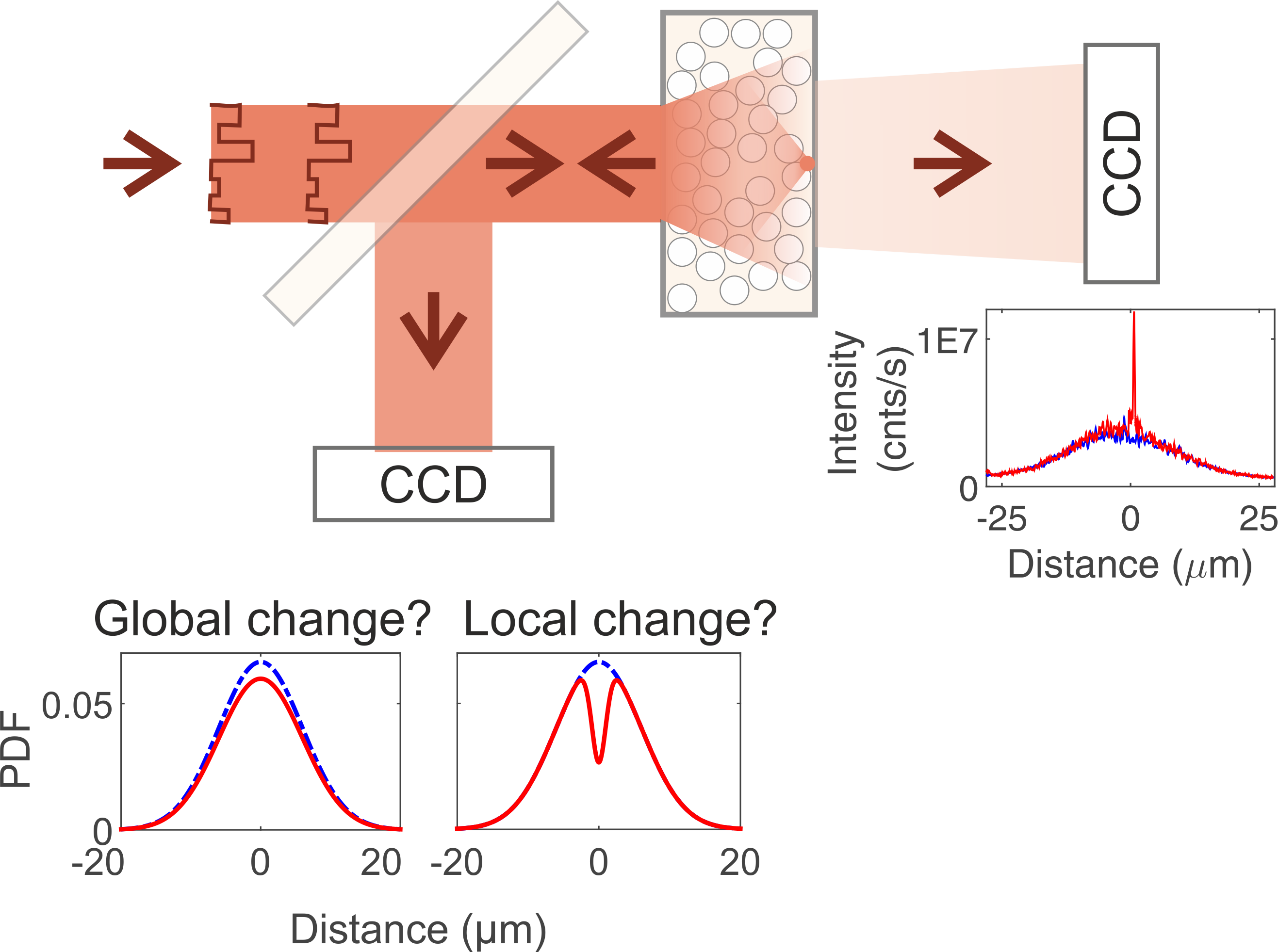}
\caption{(Color online) An illustration of the question about the type of change happens spatially on when the reflected light is decreased to enhance the transmitted light.
The wavefront-shaped light is incident on the sample to obtain an optimized focus at the back surface of the sample. 
The imaged intensity data at the back surface shows an enhanced peak (top inset).
The question is how does the reflected intensity spatially re-distribute? 
A global change (cartoon on the bottom left inset) means that there is a uniform decrease in the amplitude of the Gaussian envelope of the reflected intensity and a local change (cartoon on the bottom right inset) means there is a local dip in the Gaussian envelope.}
\label{fig:cartoon}
\end{figure}

At this point, the question arises: how does the reflected intensity change spatially? 
Is the change global, \textit{i.e.,} does the overall reflected intensity decrease uniformly, or local, \textit{i.e.,} does the intensity decrease more in the area where the transmission is enhanced? 
As illustrated in Fig.~\ref{fig:cartoon}, a global change of the reflected intensity implies that the intensity redistributed to transmission is deducted equally from all the spatial channels.
This is expected if all input spatial channels contribute equally to all the output spatial channels.
On the other hand, a local change implies that the intensity of the spatial input channels maps one-to-one with that of output spatial channels. 
The local change is expected as a result of the reflection-transmission long-range correlation predicted in Ref.~\cite{Fayard2015PRA}.

\begin{figure}[h!]
\centering
\includegraphics[width = 0.45 \textwidth]{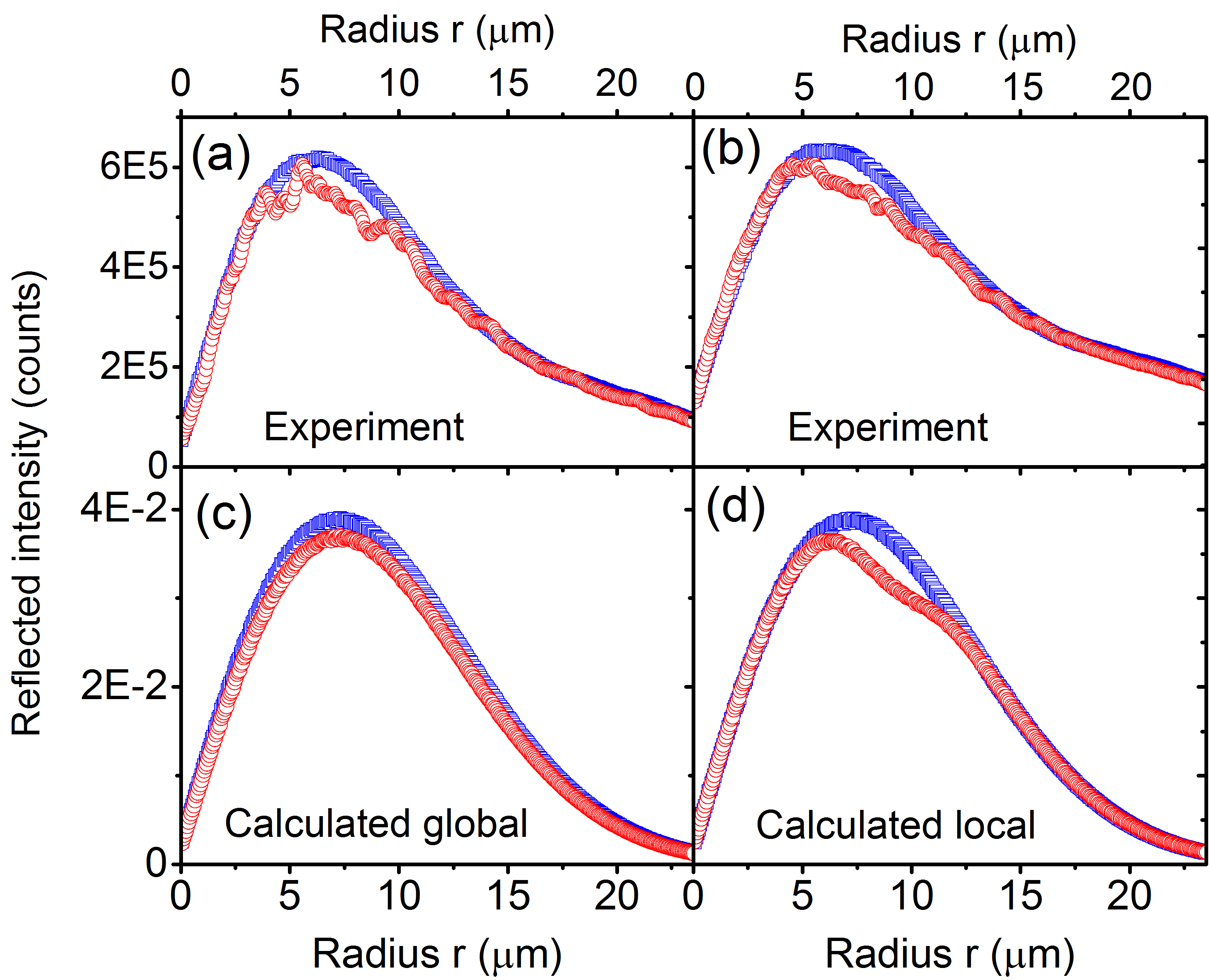}
\caption{Total reflected intensity integral along the circumference of a ring (see inset of Fig.~\ref{fig:TransmtIntensity_ring}).
In (a) and (b), the blue squares are the experimental data with unoptimized wavefront projected on the SLM and the red circles are the experimental data with the optimized wavefront.
The optimization radii are 15.2$~\mu $m and 8.4$~\mu $m in (a) and (b) respectively.
The calculated global and local change are plotted in (c) and (d) and the symbols have the same meanings as in (a) and (b).
}
\label{fig:reflIntensityRing}
\end{figure}

In order to observe the type of change, we plot the radial distribution of the reflected intensity in Figs.~\ref{fig:reflIntensityRing} (a) and (b) for optimization radii of 15.2$~\mu $m and 8.4$~\mu $m respectively.
In both optimization radii, the optimized (red circles) and unoptimized (blue squares) intensities matches quite well from $r = 0$ to about $r = $ 5$~\mu $m. 
The optimized intensity deviates asymmetrically from the unoptimized intensity between $r = $ 5$~\mu $m and 10$~\mu $m.
At $r > 10~\mu$m, both intensities become equal and decrease in the same way.

For a proper interpretation of these results, we describe the global and local change in the reflected intensity as follows.
First, we model the unoptimized reflected intensity $I_u$ with a 2D circular Gaussian function 
\begin{equation}
I_u(\pmb{\rho}) = \frac{A_1}{2\pi\sigma_1^2} \exp \left( - \left( \frac{|\pmb{\rho} - \pmb{\rho}_0|^2}{2\sigma^2_1} \right) \right) \, .
\label{eq:unoptimized}
\end{equation}
Here, $A_1$ is the amplitude, $\pmb{\rho} = (x, y)$ is the spatial coordinate, $\pmb{\rho}_0 = (x_0, y_0)$ is the coordinate of the center, and $\sigma_1$ is the standard deviation that defines the width of the function.
In order to model both the global and local change, we define a second 2D circular Gaussian function $I_o$
\begin{equation}
I_o(\pmb{\rho}) = \frac{A_2}{2\pi\sigma_2^2} \exp \left( - \left( \frac{|\pmb{\rho} - \pmb{\rho}_0 - \pmb{\rho}_c|^2}{2\sigma^2_2}  \right) \right) \, ,
\end{equation}
where $\pmb{\rho}_c = (x_c, y_c)$ is the displacement of the center of $I_o$ from $x_0$ and $y_0$ respectively and $\sigma_2$ is the standard deviation of $I_o$.
We model the optimized intensity $I_g$, in the case of the global change, as the difference between $I_u$ and $I_o$, when $\sigma_1 = \sigma_2 = \sigma_g$ and $A_1 > A_2$, to get
\begin{flalign}
I_g(\pmb{\rho}) & = 
\frac{A_2}{2\pi\sigma_g^2} \exp \left( - \left( \frac{|\pmb{\rho} - \pmb{\rho}_0 - \pmb{\rho}_c|^2}{2\sigma^2_g} \right) \right) \nonumber \\ &
- \frac{A_1}{2\pi\sigma_g^2} \exp \left( - \left( \frac{|\pmb{\rho} - \pmb{\rho}_0|^2}{2\sigma^2_g} \right) \right) \, . 
\label{eq:I_global}
\end{flalign}
To model the optimized intensity $I_l$, in case of a local change, we use the difference between $I_u$ and $I_o$, when $\sigma_1 > \sigma_2$ and $A_1 > A_2$, to get
\begin{flalign}
I_l(\pmb{\rho})  & = 
\frac{A_2}{2\pi\sigma_2^2} \exp \left( - \left( \frac{|\pmb{\rho} - \pmb{\rho}_0 - \pmb{\rho}_c|^2}{2\sigma^2_2} \right) \right)
\nonumber \\ &
- \frac{A_1}{2\pi\sigma_1^2} \exp \left( - \left( \frac{|\pmb{\rho} - \pmb{\rho}_0|^2}{2\sigma^2_1} \right) \right) \, . 
\label{eq:I_local}
\end{flalign}
Using Eqs.~\ref{eq:unoptimized}, \ref{eq:I_global} and \ref{eq:I_local}, 2D Gaussian functions were calculated for unoptimized light, and optimized light for either global or local changes, respectively. 
A projection of the generated functions onto the x-axis is shown in the bottom left and right insets of Fig.~\ref{fig:cartoon}.
Following the same procedure for the analysis of the experimental data, we obtain the radial distribution of the calculated functions by integrating the intensity along the circumference of a ring of width of $\delta r$ and inner radius of $r_r$. 

In Figs.~\ref{fig:reflIntensityRing} (c) and (d), we plot the calculated intensity versus radius for both the global and local change. 
To obtain the curves in Figs.~\ref{fig:reflIntensityRing} (c) and (d), we used these parameters: $\sigma_1 = \sigma_g = 7.5 ~ \mu $m, as obtained by fitting a Gaussian function to the unoptimized reflected intensity;  $A_1 = 1$ and $A_2 = 0.05$. 
For both global and local changes, the values of $A_1$ and $A_2$ are chosen such that the ratio of the total area under the calculated optimized and unoptimized functions is 0.95, which corresponds to a comparable enhancement on reflection in Fig.~\ref{fig:eta_total}.
Both $x_0$ and $y_0$ were chosen to be 45.5$~\mu$m, which is exactly at the center of the generated 2D Gaussian function.
The adjustable parameters are $\sigma_2 = 2.1 ~ \mu $m, and $x_c = y_c = 6.4 ~ \mu $m and we will comment on these values in the next paragraph.
For the global change shown in Figs.~\ref{fig:reflIntensityRing} (c), there is a symmetric deviation of the reflected intensity from the unoptimized light and the deviation is centered at the peak position near $7 ~ \mu $m.
At radial positions between $r = 0$ and $r = 5~ \mu $m and $r = 12~ \mu $m and $r = 26~ \mu $m, optimized light matches with unoptimized light.
These features of the calculated global change do not correspond with the features of the experimental data shown in Figs.~\ref{fig:reflIntensityRing} (a) and (b).
For the local change shown in Fig.~\ref{fig:reflIntensityRing} (d), there is interestingly an asymmetric deviation of the optimized light from the unoptimized light between $r = 5 ~ \mu $m and $r = 12 ~ \mu $m. 
This asymmetric deviation is very similar to what is observed for the two optimization radii in Figs.~\ref{fig:reflIntensityRing} (a) and (b).
Our experiment results therefore indicate that there is a local change in the reflected intensity rather than a global change. 

We now comment on the values of the adjustable parameters. 
In the experiment, $x_c \approx 0.5~ \mu$m and $y_c \approx 2.4~ \mu $m and we have used $x_c = y_c = 6.4 ~ \mu$m in Figs.~\ref{fig:reflIntensityRing} (c) and (d), in order to have a similar asymmetric deviation in the local change.
It should be noted that the global change is almost independent of $\pmb{\rho}_c$, since $|\pmb{\rho}_c| \ll |\pmb{\rho}_0|$ (see Eq.~\ref{eq:I_global}).
The discrepancy between the experiment values of ($x_c$, $y_c$) and the adjusted values might be the result of the thermal drift of the SLM, laser, mechanical stage onto which the sample is mounted and other apparatuses. 
From the experimental and adjusted values of ($x_c$, $y_c$), we estimate that the beam is displaced by approximately $7 ~ \mu $m, which is conceivable since the measurements took several days.
Consequently, due to the beam displacement of $7 ~\mu $m, the optimization area mapped onto the reflection light is almost outside the reflected intensity, which has $\sigma_1 = 7.5 ~ \mu $m.
Therefore, the large optimization radii on the transmission side does not have a significant effect on the reflection side.
This explains why $\sigma_2 = 2.1 ~ \mu $m, rather than $15.7 ~ \mu$m and $8.4 ~ \mu$m, shows in Fig.~\ref{fig:reflIntensityRing} a feature that is comparable to the feature from the experiment.
Despite these imperfections in the experiment, our data qualitatively shows that there is a local change in the optimized intensity. 

\section{Summary}
We have experimentally shown that as the optimization radius increases, the enhancement of the total transmitted intensity increases, while simultaneously the total reflected intensity decreases. 
We also find that the enhancement of the intensity outside the optimization area increases as the optimization radius increases.
From the radial intensity distribution of the reflected intensity, we find evidence that there is a local decrease in the reflected intensity rather than a global decrease.
The local decrease confirms that the transmitted and reflected intensities are spatially correlated as recently predicted by Fayard \textit{et al}~\cite{Fayard2015PRA}. 
Our results have prospects in extending the applications of wavefront shaping to increase the total transmitted intensity through the rough layer on top of the silicon absorber in a solar cell.
Our results are also interesting for multi-channel quantum secure communication~\cite{Huisman2015OptExp,Defiennee2016SciAd,Wolterink2016PRA}, where enhanced intensities are desired in multiple transmission channels; for transmitting arbitrary intensity distribution through a scattering medium~\cite{dickey2014laser}; and imaging through an opaque medium~\cite{{Vellekoop2010OptLett,Katz2014NatPhoton,Bertolotti2012Nature}}.

\acknowledgments{
We thank Duygu Akbulut, Ad Lagendijk, Ivo Vellekoop, Ravitej Uppu, and Tom Wolterink for discussions and Cock Harteveld for technical assistance. 
This project is part of the research program of the ‘Stichting voor Fundamenteel Onderzoek der Materie’ (FOM) FOM-program ‘Stirring of light!’, which is part of the ‘Nederlandse Organisatie voor Wetenschappelijk Onderzoek’ (NWO). We acknowledge NWO-Vici, DARPA, ERC 279248, and STW.
}

\appendix*
\section{Comparing wavefront shaping algorithms for large areas optimization}
For the optimization of intensity in large areas, we investigated three different wavefront shaping algorithms: the stepwise sequential algorithm, the partitioning, and the genetic algorithms. 
The details on how these algorithms work are described in Refs.~\cite{vellekoop2008OptComm,Conkey2012OptExp}. 
Firstly, the sequential algorithm modulates the phase of the segments of the SLM one by one and combines them at the end of optimization. 
Secondly, the partitioning algorithm modulates the phase of 50$\%$ of the segments simultaneously and keeps the optimized phases on the SLM.
The modulated segments are chosen randomly at each step. 
A better performance of the partitioning algorithm is expected because a larger number of segments is controlled simultaneously, and this is expected to yield a significant change in the target signal compared to the sequential algorithm.
Thirdly, the genetic algorithm begins by creating a population of random phase masks, which are ranked using the measured enhancement. 
The phase masks are combined using a weight proportional to the enhancement and then further mutated to create new phase masks. 
The new phase masks are measured and replace the low ranking members of the population. 
As the whole steps are repeated, the average enhancement of the population increases and finally converges.

\begin{figure}[h!]
\centering
\includegraphics[width = 0.4 \textwidth]{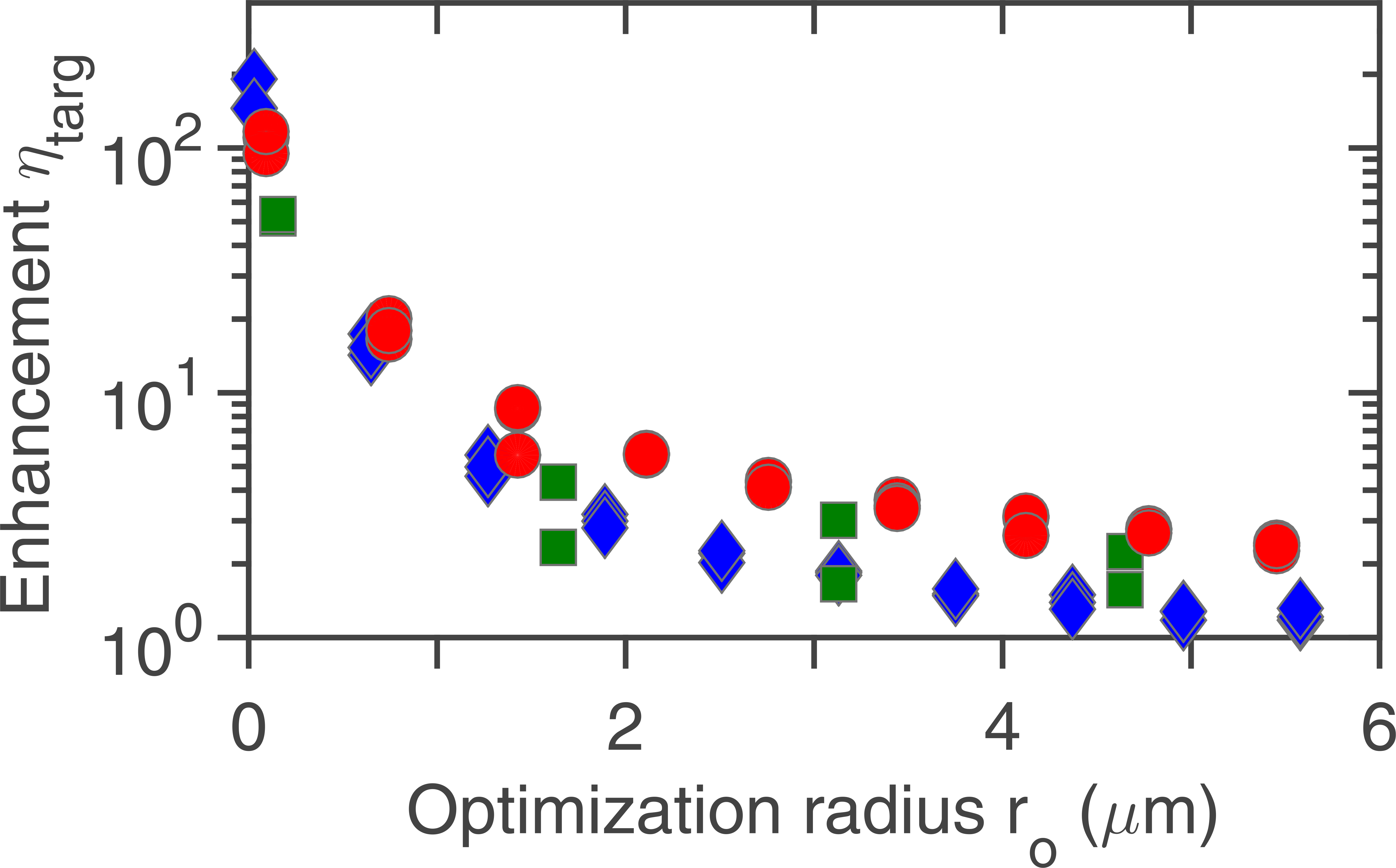}
\caption{Enhancement $\eta_{\rm{targ}}$ in optimization target versus the optimization radius $r_o$ for different algorithms.
The red circles, green squares, and blue diamond are the data point for the partitioning, genetic and stepwise sequential algorithm algorithms respectively. 
The partitioning algorithm outperforms the other algorithms.}
\label{fig:ch07_compareAlgorithms}
\end{figure}

We have performed wavefront shaping experiments to compare the enhancement of the three algorithms. 
In Fig.~\ref{fig:ch07_compareAlgorithms}, we plot the enhancement $\eta_{\rm{targ}}$ in the optimization area versus the optimization radius $r_o$.
The enhancement decreases with increasing radius for all algorithms, as is expected.
With the partitioning algorithm, the enhancement is on average approximately $85\%$ higher than for the sequential algorithm. 
The partitioning algorithm outperforms the sequential algorithm because of the larger modulation signal in the optimization area.
Furthermore, the partitioning algorithm has an enhancement that is $80\%$ higher than for the genetic algorithm. 
We expected a similar performance of the genetic algorithm and the partitioning algorithm, since a comparable number of segments is simultaneously controlled in the two algorithms.
We attribute the lower performance to the fact that the genetic algorithm requires a large number of experimental parameters, which might differ for different optimization radii. 
A further detailed study of using genetic algorithm for large areas optimization is needed. 
We have therefore chosen to use the partitioning algorithm since, it shows a better performance than the other two algorithms.

\end{document}